
\magnification=\magstep1
\tolerance 500
\rightline{IASSNS 93/51}
\rightline{TAUP 2084-93}
\rightline{September,1993}
\vskip 3 true cm
\centerline{\bf A Soluble Model for
Scattering and Decay}
\centerline{\bf in}
\centerline{\bf Quaternionic Quantum
Mechanics II: Scattering}
\centerline{L.P.
Horwitz\footnote{*}{Permanent
address: School of Physics, Raymond
and Beverly Sackler Faculty of Exact
Sciences, Tel Aviv University, Ramat
Aviv, Israel; also at Department of
Physics, Bar Ilan University, Ramat
Gan, Israel.}}
\centerline{School of Natural Sciences}
\centerline{Institute for Advanced Study}
\centerline{Princeton, N.J. 08540}
\vskip 3 true cm
\noindent
{\it Abstract.\/}  In a previous
paper, it was shown that a soluble
model can be constructed for the
description of a decaying system in
analogy to the Lee-Friedrichs model
of complex quantum theory.  It is
shown here that this model also
provides a soluble scattering
theory, and therefore constitutes a
model for a decay scattering system.
Generalized second resolvent
equations are obtained for
quaternionic scattering theory.
It is shown explicitly for this model,
in accordance with a general theorem
of Adler, that the scattering matrix
is complex subalgebra valued.
It is also shown that the method of Adler, using an
effective optical potential in the complex sector to describe the effect
of the quaternionic interactions, is equivalent to
the general method of Green's functions described here.
\vfil
\eject
\noindent
{\bf 1. Introduction}
\par Quaternionic quantum
mechanics$^1$ has recently become of
interest as a generalization of the
usual complex quantum theory with
additional intrinsic degrees of
freedom. Essentially new models for quantum field
theory have emerged from  this generalization$^1$.
 It therefore is
potentially applicable to
particle theory. These additional
degrees of freedom, with the
symmetry of the automorphisms
of the quaternion algebra, moreover,
imply that the anti-self-adjoint
operators which are generators of
groups have symmetric effective
spectra.  The generator of time
evolution, if it is absolutely
continuous on the half line, has an
effective absolutely continuous
spectrum on the whole line (this
effective spectrum is evident from
the presence of contributions from
both right and left hand cuts in the
scattering formalism developed by
Adler$^2$), so that a conjugate
symmetric ``time'' operator
exists$^3$.  The theory may,
consequently, be useful for the
description of unstable systems for
which there exists a Lyapunov
function$^4$.
\par In a previous work$^5$, it was
shown that a soluble model can be
constructed for the the description
of an unstable quaternionic quantum
mechanical system, in analogy to the
Lee-Friedrichs model$^6$ for complex
quantum theory. Following the
procedure of Wigner and
Weisskopf$^7$ to define the decay of
an unstable system, the probability
amplitude for the system to remain
in its initial (unstable) state is
(we shall use the round bracket for proper kets, i.e., normalized
vectors in the Hilbert space, and the angle bracket for the
generalized states (labels of continuous spectral representation)

$$ A(t) = ( \psi_0\vert e^{-{\tilde H}t} \vert \psi_0 ),
\eqno(1.1)$$
where
$$ {\tilde H}={\tilde H}_0 + {\tilde
V}, \eqno(1.2)$$
The vector $\psi_0$ is an eigenstate
of ${\tilde H}_0$, i.e.,
$$ {\tilde H}_0 \vert \psi_0) =
\vert \psi_0 ) iE_0,\eqno(1.3)$$
and we define the (generalized)
continuum eigenstates by
$$ \langle E \vert {\tilde H}_0
\vert f)  = iE \langle E\vert f).
\eqno(1.4)$$
\par The Laplace transform of the
amplitude $A(t)$ is defined as
$$ {\hat A}(z) = \int_0^\infty dt\
e^{izt} A(t) , \eqno(1.5)$$
where $z \in {\bf C}(1,i)$, the
complex subalgebra of the quaternion
algebra ${\bf H}$. One finds $$H(z)
{\hat A}(z) = i G(z), \eqno(1.6)$$
where H(z) is a complex-valued function of z
analytic in the upper half plane, containing
both left and right hand cuts on the real $E$ axis;
$$G(z) = G_\alpha(z) + j G_\beta(z),\eqno(1.7)$$
where $G_\alpha (z)$ is a complex-valued analytic function
of $z$ and $G_\beta(z)$ is a complex-valued analytic function
of $z^*$, as required by the quaternionic Cauchy-Riemann
relations$^5$ for left analytic $G(z)$.  The functions $H(z)$ and $G(z)$
depend only on the matrix elements $\langle E \vert {\tilde V}\vert
\psi_0)$ of the potential, and are explicitly given in ref. 5;
some of the analytic properties of $H(z)$ are discussed there as
well, making use of the full effective spectrum $(-\infty, \infty)$
of the anti-self-adjoint operator ${\tilde H}_0$.
It was  assumed, to obtain
the result $(1.7)$, that ${\tilde V}$ has
no continuum-continuum matrix
elements, i.e., $$\langle E \vert
{\tilde V} \vert E' \rangle = 0.
\eqno(1.8)$$
This restriction corresponds to that
of the Lee-Friedrichs model of
complex quantum theory.
\par In the next section, we discuss
the basic structure of formal
scattering theory in quaternionic
quantum mechanics, and define the
resolvent (Green's function) in
terms of an operator-valued Laplace
transform which exists on a certain
domain.  We then derive, in Section
3, formulas analogous to those of
the second resolvent equation
(Lippmann-Schwinger equations), and
in Section 4, use these to show that
the soluble model discussed in ref.\ 5
provides a soluble model for
scattering as well, and hence
corresponds to a decay-scattering
system$^8$. We show, explicitly, in this soluble model,
that the $T$-matrix is purely complex-valued, in accordance with a
general theorem of Adler$^1$, and furthermore show that for the general
potential problem, Adler's method for describing the solution of the
problem in the complex sector with the use of an ``optical potential,''
describing the effect of the quaternionic interactions, is equivalent to
the method of Green's functions developed here.
\bigskip
\noindent
{\bf 2. Formal Scattering Theory}
\smallskip
\par In this section we study the
formal structure of quaternionic
scattering theory. The resolvent
techniques used in scattering theory
for complex quantum mechanics do not
apply directly, since the anti-self-adjoint generator of motion in time
is not associated, in general, with any
particular complex subalgebra.
However, as in scattering theory in
the complex case, the spectral
representation of the unperturbed
Hamiltonian may be used to construct
the integral equations for the
Green's functions and the scattering
operator.  The methods we develop
here can be used to define the
spectrum of an operator, as the
complement of the resolvent set\footnote{\S}
{In the complex Hilbert space, the resolvent set for an
operator $B$ is defined
as the domain of analyticity of the operator $(z - B)^{-1}$; this is an
open set.  Its complement, a closed set, is called the spectrum$^9$.}
, and
for many parallel applications of
resolvent theory.
\par The fundamental condition for a
scattering process is the existence
of the strong limit
$$ \lim_{t \rightarrow \pm \infty}
\Vert e^{-{\tilde H}t}\psi - e^{-{\tilde H}_0 t}\phi
\Vert=0,\eqno(2.1) $$
where $\psi$ is the Heisenberg
state of the system, and $\phi$ is
an asymptotic free state.
In this norm, we may bring the
factor $e^{-{\tilde H}t}$ to the
second term. If the limit
$$ \lim_{t \rightarrow \pm \infty}
e^{{\tilde H}t}e^{-{\tilde H}_0 t}
\phi $$
exists on a dense set of the
quaternionic Hilbert space, it
defines the wave operators
$$ \Omega_\pm = \lim_{t \rightarrow
\pm \infty} e^{{\tilde H}t}e^{-{\tilde H}_0 t} \eqno(2.2)$$
with the properties
$${\tilde H} \Omega_\pm = \Omega_\pm
{\tilde H}_0, \eqno(2.3)$$
as given in ref.\ 1, Chap. 8.  A
sufficient condition for its
existence is, as for the complex
Hilbert space$^{10,11}$
$$\lim_{t \rightarrow \pm \infty}
\Vert {\tilde V} e^{-{\tilde H}_0 t}
\phi \Vert = 0 \eqno(2.4)$$
on a dense set.
\par We now define the Green's
function (the functions we define
here are the Laplace transforms of
the Green's functions defined on $t$
in ref.\ 1)
$$ G(Z) = I \int_0^\infty  dt\ e^{IZt}
e^{-{\tilde H}t} , \eqno(2.5)$$
where $Z, I  \in {\bf C}(1,I)$, the
complex subalgebra of the left
operator valued algebra isomorphic
to ${\bf H}$, constructed from the
spectral family of ${\tilde H}_0$.
The Green's function $(2.5)$ cannot
be given in closed quaternionic
operator algebraic form, since $I$
and $\tilde H$ do not, in general,
commute. The unperturbed Green's
function, however, has the form

$$\eqalign{G_0(Z) &= I \int_0^\infty dt\
e^{IZt} e^{-{\tilde H}_0t} \cr
&= -{ 1 \over Z+I{\tilde H}_0} \cr},\eqno(2.6)$$
for $Z$ formally in the upper half
plane (positive coefficient of $I$
in $Z$).

\par

In terms of the wave operators,
the scattering $S$-matrix is
$$ S= \Omega_+^\dagger \Omega_-,\eqno(2.7)$$
and the corresponding $T$-matrix,
$$ T = S - 1 \eqno(2.8)$$
can be expressed as$^{12}$

$$\eqalign{T &= \lim_{t \rightarrow -\infty} \bigl(\Omega_+^\dagger
e^{{\tilde H}t} e^{-{\tilde H}_0t} -1 \bigr)\cr
&= \lim_{t \rightarrow -\infty} e^{{\tilde H}_0 t}
\bigl(\Omega_+^\dagger - 1 \bigr) e^{-{\tilde H}_0t} ,\cr} \eqno(2.9)$$
where we have used the intertwining
property $(2.3)$ in the last step.
Denoting by $\{\langle E \vert f) \}$
the part of the spectral
representation of a vector $f$ in
the absolutely continuous spectrum
of ${\tilde H}_0$ (${\tilde H}_0$
also has a discrete eigenfunction in
the soluble model we are studying,
which we shall call $\phi$; for
example, in this case,
$\int_0^\infty dE \vert E \rangle i
\langle E \vert + \vert \phi)i (\phi
\vert \ $ represents the left algebraic operator$^{13}$
$I$), we have then that

$$\langle E \vert T\vert f) =
\lim_{t \rightarrow -\infty} e^{iEt}
\langle E \vert
\bigl(\Omega_+^\dagger -1 \bigr)
e^{-{\tilde H}_0t} \vert f).
\eqno(2.10)$$

\par   We now remark that
$$-i \lim_{\epsilon \rightarrow 0_+}
\epsilon \langle E \vert
G_0(E+I\epsilon) \vert f) = \langle
E \vert f),\eqno (2.11)$$
as can easily be seen from the
integral representation
$$-i\epsilon \langle E \vert
G_0(E+I\epsilon) \vert f) = \epsilon \int_0^\infty dt\
e^{i(E+i\epsilon)t} e^{-iEt} \langle
E \vert f), $$
which, in the limit $\epsilon
\rightarrow 0$, reduces to $\langle E
\vert f)$.  For the first term of
$(2.10)$, we use the fact that
$$ \langle E \vert \Omega_+^\dagger
\vert f) = -i \lim_{\epsilon
\rightarrow 0} \epsilon \langle E \vert
G(E+I\epsilon) \vert f). \eqno
(2.12)$$
This result follows from comparing
$$\eqalign{ \langle E \vert
\Omega_+^\dagger \vert f) &= \langle
E \vert \lim_{\epsilon \rightarrow
0} \epsilon \int_0^\infty  dt e^{-\epsilon t}
 e^{{\tilde H}_0t} e^{-{\tilde H} t} \vert f) \cr
&= \lim_{\epsilon \rightarrow 0}
\epsilon \int_0^\infty dt
e^{i(E+i\epsilon)t} \langle E \vert
e^{-{\tilde H}t}\vert f), \cr}$$
where we have used Abel's formula\footnote{*}{That is,
that $\lim_{t \rightarrow \infty}g(t) = \lim _{\epsilon \rightarrow
0} \epsilon \int_0^\infty e^{-\epsilon t} g(t) dt.$}
(assuming convergence), and
$$ \lim_{\epsilon \rightarrow 0} -i\epsilon \langle E \vert
G(E+I\epsilon)\vert f) = -i
\lim_{\epsilon \rightarrow 0}
\epsilon \langle E \vert I
\int_0^\infty dt e^{I(E+I\epsilon)t}
e^{-{\tilde H}t} \vert f),$$
which, with the properties of
$\langle E \vert$, coincides with
the previous expression, and
completes the proof of $(2.12)$.
\par We therefore have the
expression, familiar from complex
scattering theory as well,
$$\langle E \vert T \vert f) = -i
\lim_{{t \rightarrow -\infty}\atop{\epsilon \rightarrow
0_+}} e^{iEt}\epsilon  \langle E
\vert \bigl(G(E+I\epsilon) -
G_0(E+I\epsilon)\bigr) e^{-{\tilde
H}_0t} \vert f). \eqno(2.13)$$
\par Clearly, $f$ cannot contain an
eigenstate of ${\tilde H}_0$, or the
limit $t \rightarrow \infty$ will
not converge.  We must therefore
take $f$ entirely in the continuum
of ${\tilde H}_0$.
\bigskip
\noindent
{\bf 3. Second Resolvent Equations}
\smallskip
\par The relation corresponding to
$(2.13)$ in the complex quantum
theory may be analyzed with the help
of the second resolvent equations
$$\eqalign{G(z) &= G_0(z)-
G_0(z)VG(z) \cr &=G_0(z) -
G(z)VG_0(z).\cr} \eqno(3.1)$$
\par One can obtain equations analogous
to $(3.1)$ by taking the formal
Laplace transform of the derivative
of the unitary evolution operator,

$$\eqalign{ - {d \over dt} e^{-{\tilde H}t }&= ({\tilde H}_0 +
{\tilde V}) e^{-{\tilde H}t} \cr &=
e^{-{\tilde H}t} ({\tilde H}_0 +
{\tilde V}) . \cr} \eqno(3.2)$$
Multiplying from the left by $e^{IZt}$ and
integrating between $(0,\infty)$,
one obtains from the first of
$(3.2)$ [the first term on the left hand side of
these equations is due to integration by parts]
$$1+ ZG(Z)= -I{\tilde H}_0 G(Z) +
\int_0^\infty dt\ e^{IZt}{\tilde V}
 e^{-{\tilde H}t}  \eqno(3.3)$$
and from the second of $(3.2)$,
$$1+ZG(Z) = -IG(Z)({\tilde H}_0 +
{\tilde V}). \eqno(3.4)$$
\par We now make use of the symplectic
decomposition available for
quaternion linear operators$^{1,13}$
to define
$$ {\tilde V} = V_\alpha + JV_\beta ,
\eqno(3.5)$$
where $V_\alpha$ and $V_\beta$
are ${\bf C}(1,I)$ valued
operators, and $J$ is the second
generator of the algebra, isomorphic
to ${\bf H}$, spanned over the reals
by the left multiplying operators
$\{1,I,J,K\}$.  From $(3.3)$, we then obtain
$$ 1+ ZG(Z) = -I{\tilde H}_0 G(Z) -IV_\alpha G(Z) -JIG(-Z^*).
\eqno(3.6)$$
Substituting the decomposition
$$G(Z) = G_\alpha (Z) + J
G_\beta(Z), \eqno(3.7)$$
we obtain
$$1 + ZG_\alpha(Z) = -I{\tilde H}_0 G_\alpha(Z) -IV_\alpha G_\alpha(Z)
-IV_\beta^*G_\beta(-Z^*) \eqno(3.8)$$
and
$$ Z^* G_\beta(Z) = -I{\tilde H}_0 G_\beta(Z) + IV_\alpha^* G_\beta(Z)
-IV_\beta G_\alpha(-Z^*);\eqno(3.9) $$
we use the asterisk here to represent the involution of the left
quaternionic algebra as well as its (isomorphic) use as quaternion
conjugation of the right algebra.  For formally complex-valued
operators $B_\alpha$, we have that $B_\alpha^* \equiv JB_\alpha J^*$.
\par With the definition $(2.6)$, one then obtains from $(3.8)$ and
$(3.9)$
the two equations (see ref. 1, Chap. 7, Sec. 7.2 and Chap. 8, and
ref. 2 for related results)
$$G_\alpha(Z) = G_0(Z) +
G_0(Z)IV_\alpha G_\alpha(Z) +
G_0(Z)V_\beta^* IG_\beta(-Z^*)
\eqno(3.10)$$
and
$$G_\beta(Z) = -G_0(Z)^*IV_\alpha^*
G_\beta(Z) + G_0(Z)^*IV_\beta
G_\alpha(-Z^*). \eqno(3.11)$$
Similary, from $(3.4)$, substituting $(3.5)$ and $(3.7)$, we obtain
$$\eqalign{ 1 + ZG_\alpha(Z) &= -IG_\alpha(Z)({\tilde H}_0 + V_\alpha)
+ I G_\beta(Z)^*V_\beta \cr
Z^* G_\beta(Z) &= IG_\beta(Z)({\tilde H}_0 + V_\alpha) +
IG_\alpha(Z)^* V_\beta \cr}. \eqno(3.12)$$
It then follows, again using $(2.6)$, that
$$G_\alpha(Z) = G_0(Z) +IG_\alpha(Z)
V_\alpha G_0(Z) -
IG_\beta(Z)^*V_\beta G_0(Z)
\eqno(3.13)$$
and
$$G_\beta(Z)= IG_\alpha(Z)^*V_\beta
G_0(-Z^*) + IG_\beta(Z)V_\alpha G_0(-Z^*). \eqno(3.14)$$
The formulas $(3.10)$, $(3.11)$, $(3.13)$ and $(3.14)$ are the
generalized second resolvent
formulas for quaternionic Green's
functions.
\bigskip
\noindent
{\bf 4. Solubility of the Model}
\smallskip
\par  In this section we apply the
generalized second resolvent
equations to the soluble model
discussed in ref. 5.  The model is described by
Eqs. $(1.2)-(1.4)$, with the condition $(1.8)$.
The matrix element $\langle \psi_0 \vert
 {\tilde V}\vert \psi_0 \rangle$,
 included in the analysis
of the decay system carried out in ref. 5, does not
play a direct role in the structure of
the equations for the scattering amplitude,
but the Green's function $G(Z)$ appearing in the formula
$(2.13)$ depends implicitly on this part of the potential as well$^5$.
\par To display explicitly the
property that the continuum-continuum matrix elements of the
interacting Green's functions depend
only on the discrete-discrete matrix
elements (with transition amplitudes
provided by the symplectic
components of the potential) in the
soluble model we are considering, we
use $(3.13)$ and $(3.14)$ in $(3.10)$ and $(3.11)$
to obtain
$$\eqalign{G_\alpha (Z) &= G_0(Z) +
G_0(Z)\bigl[IV_\alpha -V_\alpha
G_\alpha(Z)V_\alpha \cr &+ V_\alpha
G_\beta(Z)^*V_\beta -V_\beta^*
G_\beta(-Z^*)V_\alpha
- V_\beta^* G_\alpha(-Z^*)^*V_\beta \bigr]G_0(Z), \cr} \eqno(4.1)$$
and
$$\eqalign{G_\beta(Z) &=
G_0(Z)^*\bigl[ IV_\beta +V_\alpha^*
G_\beta(Z) V_\alpha + V_\alpha ^*
G_\alpha(Z)^* V_\beta \cr &- V_\beta
G_\alpha(-Z^*) V_\alpha + V_\beta
G_\beta(-Z^*)^* V_\beta \bigr] G_0(-Z^*). \cr} \eqno(4.2)$$
\par We now return to Eq.$(2.13)$
for the $T$-matrix; recalling that
the vector $f$ must be in the
continuous spectrum of ${\tilde
H}_0$ if the limits are to be well-defined, we see that we are
interested in the continuum-continuum matrix elements of $$G(Z)
- G_0(Z) = G_\alpha(Z) - G_0(Z) +
JG_\beta(Z). \eqno(4.3)$$
It is clear from the structure of
the potential $(1.8)$, and the form of
$(4.1)$ and $(4.2)$ of the
symplectic components of the Green's
function, that the continuum-continuum matrix elements entering
the definition of the $T$ matrix
are, in fact, expressed in terms of
the discrete-discrete matrix
elements of the Green's function
only.  Since a closed formula can be
obtained for the discrete-discrete
matrix element of the Green's
function\footnote\ddag{A closed
formula is given for ${\hat A}(z)$
in ref. 5; from the definition
$(2.5)$ of the Green's function and
of the amplitude ${\hat A}(z)$ in Eqs.$(2.1)$ and
$(3.5)$ of ref. 5  it follows that $$\langle \psi_0 \vert
G(Z)\vert \psi_0 \rangle = i {\hat A}(z).$$}
 in this model, this
result completes the proof that the
scattering theory for this model is
exactly soluble as well.
\par  In the following, we shall
give the explicit form of the $T$-matrix,

$$ \langle E \vert T \vert f) = -i
\lim_{{t \rightarrow -\infty}\atop
{\epsilon \rightarrow 0_+}} \epsilon
e^{iEt} \langle E \vert
(G_\alpha(E+I\epsilon) -G_0(E+I\epsilon) +
JG_\beta(E+I\epsilon))e^{-{\tilde H}_0t} \vert f ). \eqno(4.4)$$

For the contribution of $G_\alpha -G_0$ to $(4.4)$,
we see from $(4.1)$ that the first operator
valued factor (leftmost) in the
matrix element is
$G_0(E+I\epsilon)$; under the limit
$\epsilon \rightarrow 0_+$, this
factor, according to $(2.11)$, with
the pre-factor $-i\epsilon$, has the
effect of multiplication by unity.  The contribution
of the $JG_\beta$ term contains, as seen from $(4.2)$, for
the first operator valued factor,
$JG_0(E+I\epsilon)^* =
G_0(E+I\epsilon)J$.  Hence, applying
$(2.11)$ again, the result  (after
the limit $\epsilon \rightarrow
0_+$) is

$$\langle E \vert J\vert f') =
j\langle E \vert f'), \eqno(4.5)$$
where $f'$ is the vector generated
from $f$ by the action of the
remaining operators.
\par The last operator valued factor
in the two types of terms in the
matrix element, i.e.,
$G_0(E+I\epsilon)$ and $G_0(-E+I\epsilon)$, combine with the
prefactor $e^{iEt}$ and the unitary
operator $e^{-{\tilde H}_0t}$, in
the limit $t \rightarrow -\infty$
to form delta functions. We now
study these contributions.  Since
$f$ is in the continuous spectrum of
${\tilde H}_0$, we may represent it by

 $$ f
= \int_0^\infty \vert E'
 \rangle  \langle E' \vert f)\
dE'. \eqno(4.6)$$
Consider, for
the first of these types, the
expression
$$ e^{IEt} G_0(E+I\epsilon)e^{-{\tilde H}_0t} \vert E'
 \rangle.\eqno(4.7)$$
Using the definition $(2.6)$ for
$G_0$, one obtains
$$ \lim_{t \rightarrow -\infty}\vert
E' \rangle{e^{-i(E'-E)t} \over {(E'-E) -i\epsilon}}=
 \vert E'\rangle 2\pi i \delta(E-E').\eqno(4.8)$$
\par For the second type, there is a
factor $j$, so that we must evaluate, in place of $(4.7)$, the limit
$$ e^{-IEt} G_0(-E+I\epsilon)e^{-{\tilde H}_0t}\vert E' \rangle ;
\eqno(4.7')$$
 the corresponding
result is
$$  \lim_{t
\rightarrow -\infty}\vert E'\rangle{ e^{-i(E+E')t}
\over {E+E' -i\epsilon}}= \vert E'\rangle 2\pi i
\delta(E+E'). \eqno(4.9)$$
Since $E,\  E'>0$, this second term can give no
 contribution.  The $T$-matrix is therefore entirely
 complex-valued, relative to the
ray convention we have chosen, in accordance with a
theorem of Adler$^1$, stating that
the $S$-matrix necessarily has values entirely in the
complex subalgebra ${\bf C}(1,i)$.

\par Inserting these results into
the formula $(4.4)$ for the $T$-matrix, we have
that
$$\langle E \vert T\vert f) = \int_0^\infty dE\ \langle E \vert T \vert E'
\rangle \langle E'\vert f), \eqno(4.10)$$
where $\langle E \vert T \vert E'
\rangle$ is complex-valued:
$$ \langle E \vert T \vert E' \rangle = 2 \pi i \delta (E-E')
\langle E\vert {\cal O} \vert E' \rangle, \eqno(4.11)$$
where, from $(4.1)$ and $(4.2)$
(the $IV_\alpha$ term of $G_\alpha -
G_0$ and the $IV_\beta$ term of
$(4.2)$ do not contribute, in this
model, to the continuum-continuum
matrix elements),
$$\eqalign{ {\cal O} &= -V_\alpha
G_\alpha(E+I\epsilon)V_\alpha +
V_\alpha G_\beta(E+I\epsilon)^*
V_\beta \cr
&-V_\beta^*G_\beta(-E+I\epsilon)V_\alpha -
V_\beta^*G_\alpha(-E+I\epsilon)^*V_\beta. \cr } \eqno(4.12)$$

One could, alternatively use
 here (in place of $(4.6)$) the
spectral representation for $\vert f)$
which makes explicit the symmetric
spectrum$^3$ of ${\tilde H}_0$, i.e.,
$$ \vert f) = \int_{-\infty}^\infty dE\ \|E\rangle \ {}_c\!\langle E \|f),$$
where $\| E \rangle$ admits both positive and negative values of $E$
with the identification
 $\| - |E|\rangle = |E\rangle j$.
  The
subscript $c$ on the bra
${}_c\!\langle E \| $
indicates that the complex scalar
product must be taken$^{13}$.  The equivalence between
$(4.6)$ and this expression can be seen by noting$^3$ that
${}_c\!\langle E \vert f) = f_\alpha(E)$, and therefore
 $$\eqalign{ \int_{-\infty}^\infty
dE \ \| E \rangle
\langle E \|f) &= \int _0^\infty dE\ \biggl(\vert E \rangle
\langle E\vert f)_c + \vert E \rangle j (-j\langle E \vert f)_c)
\biggr) \cr
&= \int_0^\infty dE\ \biggl( \vert E \rangle f_\alpha(E) + \vert E
\rangle j f_\beta (E).\cr }$$
The limit  of $(4.7)$ in
the $E' <0$ interval contains  a sign
change (due to the factor $j$) in the
 eigenvalue of the operator $I$ (with the
parameter $E'$ in the coefficients positive).  Hence only the sign of
$i$ in both numerator and denominator of the
left hand side  of $(4.8)$ is changed; the sign of the
right hand side is therefore also changed, consistent with simply
multiplying both sides of $(4.8)$ by $j$ on the right.
Similarly, in this interval, $(4.7')$ leads to $(4.9)$ with the
sign of $i$ changed on both left and right sides.
  The construction of the $E' <0$ extension therefore
commutes with the limiting process, and leads to the same conclusion.
\bigskip
\noindent
{\bf 5. The Optical Potential}
\smallskip
\par Since the $T$-matrix is complex-valued and acts separately
in the complex and quaternionic sectors, the formulation of Adler$^1$
for scattering in the complex sector with the use of an effective
potential, called the ``optical potential,'' to provide the effect of
perturbation due to the interactions of the quaternionic sector should,
in fact, coincide with the results we have obtained from the general
theory of Green's functions. To obtain this equivalence, we return to
Eqs.$(3.8)$, $(3.9)$ and $(3.12)$.
 We define the Green's function associated with the complex part of the
full Hamiltonian by
$$ G_c(Z)  = - {1 \over {Z+I{\tilde H}_0 +IV_\alpha}}. \eqno(5.1)$$
Grouping terms in a different way, it then follows that,
 from $(3.8)$ and $(3.9)$,
$$ \eqalign{G_\alpha(Z) &= G_c(Z) + G_c(Z) V_\beta^* I G_\beta(-Z^*) \cr
G_\beta(Z) &= G_c(Z)^*IV_\beta G_\alpha(-Z^*) \cr}\eqno(5.2)$$
and, from $(3.12)$, we obtain
$$\eqalign{G_\alpha(Z) &= G_c(Z)- IG_\beta(Z)^*V_\beta G_c(Z) \cr
G_\beta(Z) &= IG_\alpha(Z)^*V_\beta G_c(-Z^*) . \cr}\eqno(5.3)$$
Substituting the quaternionic part $G_\beta(Z)$ of the Green's function
into the first of each of these equations, one obtains
$$ G_\alpha(Z) = G_c(Z) - G_c(Z)V_\beta^*G_c(-Z^*)^*V_\beta
G_\alpha(Z) \eqno(5.4)$$
and
$$G_\alpha(Z) = G_c(Z)- G_\alpha(Z) V_\beta^*(Z) G_c(-Z^*)^*
V_\beta G_c(Z). \eqno(5.5)$$
Making the identifications with the notation of Adler$^1$, for which the
complex part of the Hamiltonian $H_\alpha$,the unperturbed Hamiltonian
${\tilde H}_0$
 and the complex part of the
potential $V_\alpha$ are related to Hermitian operators by
$$ \eqalign{H_\alpha &= IH_1, \cr
{\tilde H}_0 &= IH_0, \cr
V_\alpha &= IV_1, \cr} \eqno(5.6)$$
we see that the conjugated Green's function appearing in $(5.4)$ and
$(5.5)$
$$ G_c(-Z^*)^* = {1 \over {Z - I{\tilde H}_0 -IV_\alpha}}, \eqno(5.7)$$
an analytic function in the upper half plane, has the form
$$\eqalign{G_c(-Z^*)^* &= {1 \over {Z + H_0 + V_1}} \cr
            &= {1 \over {Z + H_1}}, \cr} \eqno(5.8)$$
so that we recognize that
$$ V_{opt} (Z) = V_\beta^* G_c(-Z^*)^* V_\beta \eqno(5.9)$$
is the analytic function in the upper half plane with boundary value
$V_{opt}(E)$ on the real axis defined by Adler$^1$.
  If $H_1$ has positive spectrum, there is
no discontinuity as one approaches the real line.
\par Rewriting $(5.4)$ and $(5.5)$, we obtain
$$ G_\alpha(Z) = G_c(Z) - G_c(Z)V_{opt}(Z)G_\alpha(Z) \eqno(5.10)$$
and
$$G_\alpha(Z) = G_c(Z)- G_\alpha(Z) V_{opt}(Z) G_c(Z). \eqno(5.11)$$
\par These are the second resolvent equations for a complex quantum
mechanical scattering problem for which the ``unperturbed'' Green's
function
corresponds to the scattering problem with the complex part of the
quaternionic potential; the complete problem, described by
$G_\alpha(Z)$(since the part proportional to
$j$ in $(4.4)$ vanishes in the scattering limit),
 carries the perturbation induced by the interactions of
the quaternionic sector through the optical potential.  This corresponds
to the method used by Adler$^1$ to solve the scattering problem in the
complex sector, and therefore the solutions for the $T$-matrix coincide.
\par From $(5.10)$ and $(5.11)$ we see that
$$ G_\alpha (Z) = G_c(Z) \bigl(G_\alpha(Z)^{-1} - V_{opt}(Z) \bigr)
G_\alpha(Z) \eqno(5.12)$$
and
$$ G_\alpha(Z) = G_\alpha(Z) \bigl(G_\alpha(Z)^{-1} -V_{opt}(Z) \bigr)
G_c(Z), \eqno(5.13)$$
so that the effective Hamiltonion for the problem is the complex part plus
the (energy dependent) optical potential.

 \bigskip
\noindent
{\bf 6. Conclusions}
\smallskip
\par In a previous work$^{5}$ a
soluble model for the description of
a decaying system was discussed.  In
this model, the unperturbed
Hamiltonian has an absolutely
continuous spectrum on $[0,\infty)$
(and hence, an effective absolutely
continuous spectrum$^3$ on $(-\infty,\infty)$)and a discrete
eigenstate; the perturbing potential
has no continuum-continuum matrix
elements.  This property makes it
possible to obtain a closed formula
for the exact computation of the
diagonal matrix element of the full
Green's function in the discrete
eigenstate of the unperturbed
Hamiltonian (this result can be
easily generalized to a finite set
of discrete eigenstates, as for the
complex case treated, e.g., in
ref.\ 8). We have shown here that this
model also provides an exactly
soluble model in the scattering
channel.  \par From the fundamental
strong convergence relation of
scattering theory, a formula for the
scattering $T$-matrix was obtained
in terms of the Green's functions of
the theory.  The full quaternionic
Green's function does not have a
simple quaternionic operator form,
but from its definition as a formal
integral (convergent on some
domain), the quaternionic analog of
the second resolvent equations were
developed. It was then shown that
the $T$-matrix, expressed in term of
the operator $G-G_0 = G_\alpha -G_0
+JG_\beta$ reduces to a form in
which the Green's functions appear
between two factors consisting of
the complex or quaternionic parts of
the potential.  Since the potential,
in this model, connects the
unperturbed continuous spectrum only
to the discrete, the $T$-matrix then
becomes exactly solvable in terms of
the known solutions$^5$ for the
discrete matrix element of the
Green's function.  Hence, the model
provides an exactly soluble
scattering theory as well, and
serves as a decay-scattering system$^8$.
\par The result for the $T$ matrix
is purely complex-valued, in
accordance with a theorem of
Adler$^1$. We have shown, moreover, that in the framework of the general
scattering theory, the Green's function that generates the $T$-matrix
is precisely the Green's function for the complex part of the problem
perturbed by the ``optical potential'' defined by Adler$^1$.  In his
formulation of the scattering problem in the complex sector with the
effective perturbation due to the quaternionic interactions, the optical
potential plays the same role.  Hence the $T$-matrix defined by Adler
with the optical potential in the complex sector coincides (as it must,
since the $T$-matrix is purely complex valued, and acts in each sector
separately)
with the
scattering matrix we have obtained in the framework of the general
theory of quaternionic Green's functions.
\bigskip
\noindent
{\it Acknowlegements.\/} The author is grateful to C.
Piron  for
invaluable discussions of quantum
physics and
for his hospitality at the
Department of Theoretical Physics at
the University of Geneva, where the author's work was partially
supported by the Swiss National Fund.
  He is also indebted to J.-P. Marchand,
from whom he learned, many years ago,
some of the techniques of scattering
theory used here.  He also wishes to
thank S.L. Adler for many
discussions of quaternionic quantum
mechanics, for his very helpful comments on this
paper, and for his hospitality
at the Institute for Advanced Study,
where this work was supported in part by the
Monell Foundation.
\vfill
\eject
\noindent
{\bf References}
\smallskip
\frenchspacing
\item{1.} S.L. Adler, {\it
Quaternionic Quantum Mechanics and Quantum Fields},
Oxford University Press, Oxford
(1994).
\item{2.} S.L. Adler, Phys. Rev. D
{\bf 37}, 3654 (1988).
\item{3.} L.P. Horwitz,
 Jour. Math. Phys.{\bf 34}, 3405 (1993).
\item{4.} B. Misra, Proc. Nat. Aca.
{\bf 75},1627 (1978); B. Misra, I.
Prigogine and M. Courbage, Proc.
Nat. Aca. {\bf 76}, 4768 (1979).
\item{5.} L.P. Horwitz, `` A
Soluble Model for Scattering and
Decay in Quaternionic Quantum
Mechanics I: Decay,'' Inst. for Adv.
Study preprint IASSNS 92/75, Tel
Aviv University preprint TAUP 2073-93.
\item{6.} K.O. Friedrichs, Comm.
Pure Appl. Math. {\bf 1}, 361
(1950); T.D. Lee, Phys. Rev. {\bf
95}, 1329 (1956).
\item{7.} V.F. Weisskopf and E.P.
Wigner, Zeits. f. Phys. {\bf 63}, 54
(1930);{\bf 65}, 18 (1930).
\item{8.}L.P. Horwitz and J.-P.
Marchand, Rocky Mtn. Jour. Phys.
{\bf 1}, 225 (1973).
\item{9.} See, for exampple, M. Reed and B. Simon,{\it Methods of Modern
Mathematical Physics I.  Functional Analysis}, Academic Press, New York
(1980).
\item{10.} W.O. Amrein, J.M. Jauch and
K.M. Sinha,{\it Scattering Theory in Quantum Mechanics},
W.A. Benjamin Advanced Book Program, Reading, Mass.(1977).
\item{11.} R.G. Newton,{\it Scattering Theory of Waves and
Particles}, McGraw Hill, New York (1966).
\item{12.} We use here methods of
the type discussed by J.-P.
Marchand,{\it Lectures in
Theoretical Physics}, Vol. X-A, ed.
A.O. Barut and W. Brittin, p. 49,
Gordon and Breach, N.Y. (1968).
\item{13.} L.P. Horwitz and L.C.
Biedenharn, Ann. Phys.{\bf 157},432
(1984).
\vfill
\eject
\end
\bye